\documentclass[sigconf]{acmart}
\acmConference[none]{ }{ }{ }
\acmBooktitle{}

\makeatletter
\@ACM@nonacmtrue
\makeatother
\usepackage{listings}
\usepackage{algorithm}
\usepackage{algpseudocode}
\usepackage{makecell}
\usepackage{listings}
\usepackage{xcolor}

\usepackage{listings}
\usepackage{xcolor}
\usepackage{enumerate}
\usepackage{enumitem}
\usepackage{listings}
\usepackage{xcolor}
\usepackage{tabularx}
\usepackage{graphicx}
\usepackage{lmodern}
\usepackage{subcaption}
\usepackage{pifont} 
\usepackage{booktabs}
\newcolumntype{L}[1]{>{\raggedright\arraybackslash}p{#1}} 
\newcolumntype{Y}{>{\raggedright\arraybackslash}X}

\usepackage[compact]{titlesec}
\usepackage[belowskip=0pt,aboveskip=2pt,labelfont=bf]{caption}

\renewcommand\footnotetextcopyrightpermission[1]{} 
\setcopyright{none}
\acmSubmissionID{123-A56-BU3}

\newcommand{\signpost}[1]{\noindent \textbf{#1.}}

\begin{document}

\title{Towards NWDAF-enabled Analytics and Closed-Loop Automation in 5G Networks}

\author{
    Fatemeh Shafiei Ardestani, 
    Niloy Saha,
    Noura Limam,
    Raouf Boutaba
    }
\affiliation{%
  \institution{Cheriton School of Computer Science}
  \institution{University of Waterloo}
  \city{Waterloo}
  \state{Ontario}
  \country{Canada}
}
\email {{f2shafie, n6saha, noura.limam, rboutaba }@uwaterloo.ca}

\begin{abstract}
The fifth generation of cellular technology (5G) delivers faster speeds, lower latency, and improved network service alongside support for a large number of users and a diverse range of verticals. This brings increased complexity to network control and management, making closed-loop automation essential. In response, the 3rd Generation Partnership Project (3GPP) introduced the Network Data Analytics Function (NWDAF) to streamline network monitoring by collecting, analyzing, and providing insights from network data. While prior research has focused mainly on isolated applications of machine learning within NWDAF, critical aspects such as standardized data collection, analytics integration in closed-loop automation, and end-to-end system evaluation have received limited attention. This work addresses existing gaps by presenting a practical implementation of NWDAF and its integration with leading open-source 5G core network solutions. We develop a 3GPP-compliant User Plane Function (UPF) event exposure service for real-time data collection and an ML model provisioning service integrated with MLflow to support end-to-end machine learning lifecycle management. Additionally, we enhance the Session Management Function (SMF) to consume NWDAF analytics and respond accordingly. Our evaluation demonstrates the solution's scalability, resource efficiency, and effectiveness in enabling closed-loop security management in 5G networks.

\end{abstract}
\maketitle
\section{Introduction} 
\label{sec:introduction}

The fifth generation of mobile networks (5G) builds on Software-Defined Networking (SDN) and Network Function Virtualization (NFV) to support a wide array of use cases beyond traditional mobile broadband. Unlike previous generations, 5G targets not only consumer applications but also mission-critical services and industrial systems, delivering faster speeds, lower latency, and improved network service. 

While this versatility unlocks transformative capabilities, it also introduces complexity in managing network performance, ensuring service-level agreements, and maintaining robust security. To meet these challenges, the 3rd Generation Partnership Project (3GPP) introduced the Network Data Analytics Function (NWDAF) as a native component of the 5G core functions to facilitate network monitoring and intelligent network operations and management by collecting data from various network functions, applying statistical or machine learning-based methods, and generating actionable analytics.

While network operators can benefit from NWDAF-generated analytics for informed decision-making, these can also be \emph{consumed} by other network functions for the closed-loop automation of network operations. As illustrated in Figure~\ref{fig:system-overview}, NWDAF enables a feedback mechanism in which network data is continuously gathered, analyzed, and acted upon.
\begin{figure}[!ht]
    \centering
    \includegraphics[width=0.9\columnwidth]{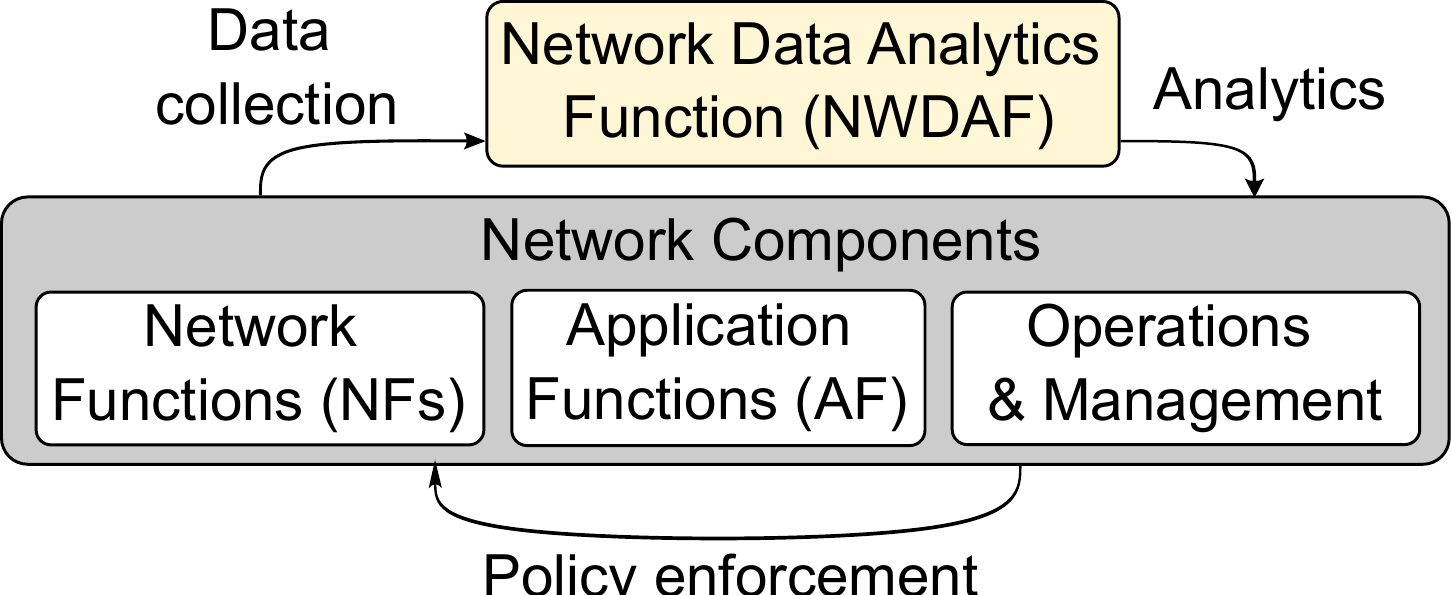}
    \caption{NWDAF-enabled closed loop network automation}
    \label{fig:closed-loop}
    \Description{}
    
\end{figure}

Despite NWDAF's potential to enable network automation and zero-touch management, most existing open-source implementations focus narrowly on applying machine learning to generate analytics and derive insights, with limited attention given to how these insights can be operationalized for real-time decision-making or performance optimization.

Another underdeveloped area is the standardized acquisition of network data, which is often handled through non-standard approaches. For example, prior work, such as Monarch \cite{monarch-tnsm24}, proposed a scalable 5G network monitoring architecture using cloud-native telemetry tools like Prometheus and Thanos. While effective for key performance metric collection, Monarch lacked native integration with 3GPP-defined control mechanisms such as NWDAF or standardized interfaces like the User Plane Function (UPF) Event Exposure Service. Although the 3GPP defines the UPF EES to enable direct, real-time data collection from the User Plane Function, this feature is largely absent from existing NWDAF implementations. Instead, data collection is often conducted through proprietary or ad hoc mechanisms, typically involving the Session Management Function (SMF), which reduces reliability, interoperability, and limits scalability.

Furthermore, current research rarely addresses the operational overhead of deploying NWDAF. There is a notable lack of quantitative evaluations examining the resource costs associated with data collection, analytics generation, and reporting, or the effectiveness of NWDAF in closed-loop automation.

This work addresses existing gaps by presenting a practical implementation of NWDAF and its integration with widely used open-source solutions (OpenAirInterface and Open5GS), coupled with a comprehensive evaluation of its efficiency and effectiveness. We present the first implementation of the UPF Event Exposure Service, enabling standardized real-time data collection from the UPF. Furthermore, we are the first to provide a proof-of-concept implementation of closed-loop automation of threat detection and mitigation involving UPF, NWDAF, and an extended version of the Session Management Function. We make the source code publicly available to facilitate further research in closed-loop automation and zero-touch management of 5G networks\footnote{https://github.com/fatemeshafiee/closed-loop-nwdaf}. Our key contributions are as follows: 

\begin{itemize}[leftmargin=*]

\item We implement the UPF Event Exposure Service (EES), allowing for the standardized collection of traffic data and monitoring of user communication behavior. 

\item We extend the Session Management Function to subscribe to NWDAF reports on abnormal User Equipment (UE) behavior and automatically terminate the Packet Data Unit (PDU) sessions of anomalous users. 

\item We develop an ML Model Provisioning Service integrated with MLflow, enabling lifecycle management of machine learning models. 

\item We systematically evaluate the performance of our system. Our results show that the UPF Event Exposure Service introduces minimal overhead, NWDAF components remain lightweight except for ML inference, and the closed-loop pipeline can detect and mitigate attacks with low latency.

\end{itemize}

The remainder of this paper is structured as follows. Section~\ref{sec:background} reviews related work and provides essential background on NWDAF. Section~\ref{sec:system} presents our system architecture, detailing enhancements to support closed-loop security automation. Section~\ref{sec:implementation} describes the implementation of each system component. Section~\ref{sec:evaluation} demonstrates the operational workflow and evaluates the system against multiple metrics. Section~\ref{sec:conclusion} concludes with key takeaways and discusses future directions for enhancing NWDAF in practical security scenarios.

\section{Background and Related Work }
\label{sec:background}
\subsection{A Primer on NWDAF and EES}
The 5G core network adopts a service-based architecture (SBA) as defined in 3GPP Release 15 and subsequent releases. In this architecture, network functions (NFs) communicate with each other through standardized interfaces. This work focuses on three key NFs: the User Plane Function, the SMF, and the Network Data Analytics Function. The UPF is responsible for handling data traffic between the UE and the data network by establishing a PDU session \cite{3gpp_ts_23_501}. In contrast, the SMF manages these PDU sessions by establishing, modifying, and releasing them via the N4 interface \cite{3gpp_ts_23_502}.

\smallskip

\signpost{NWDAF overview} The Network Data Analytics Function is a core component of the 5G network, designed to collect data from various network entities, analyze it, and deliver analytics to authorized consumers. Both 5G core network functions and Operations, Administration, and Maintenance (OAM) systems can supply data to NWDAF and utilize its analytics services. To provide these services, NWDAF can implement one or both of the following logical functions, as defined in 3GPP Technical Specification 23.288 \cite{3gpp_ts_23_288}: the Analytics Logical Function (AnLF) and the Model Training Logical Function (MTLF). The AnLF performs inference tasks, generates statistical or predictive analytics, and exposes these insights as services. The MTLF, on the other hand, is responsible for training machine learning models and offering model training services to support the AnLF or external consumers.

\smallskip

\signpost{Data collection via EES} NWDAF can collect data by subscribing to the EES offered by various network functions. In this context, NWDAF acts as a consumer, specifying subscription parameters such as the desired reporting frequency (e.g., periodic updates) and the maximum number of notifications it wishes to receive. Upon successful subscription, NWDAF receives event notifications in accordance with the parameters defined in its request.

Among the available event exposure services, the UPF event exposure service is particularly critical for NWDAF, as it enables access to detailed information on UE communications. Through this service, NWDAF can subscribe to a range of events, including Quality of Service (QoS) monitoring, user data usage measurements, user data usage trends, and Time-Sensitive Communication (TSC) management information \cite{3gpp_ts_23_502}.

The QoS monitoring event provides performance metrics at the QoS flow level, such as uplink, downlink, and round-trip packet delays. User data usage measurements include volume and throughput metrics, as well as application-level usage data. Additionally, user data usage trends offer statistical insights into throughput patterns over time.

Subscribers can customize the type and scope of data they wish to receive using various parameters defined in 3GPP specifications \cite{3gpp_ts_29_564}. For example, measurement granularity can be set at the level of individual data flows, PDU sessions, or applications. Subscriptions can also define the reporting frequency, the total number of expected notifications, and the expiration time of the subscription. Furthermore, data can be filtered based on specific attributes such as Data Network Name (DNN), Single Network Slice Selection Assistance Information (S-NSSAI), application identifiers, or traffic filtering criteria.

\smallskip

\signpost{Analytics and closed-loop automation} NWDAF is envisioned to offer a broad range of analytics services to authorized consumers within the 5G core network. These services can be accessed primarily in two ways: Consumers may either (1)~subscribe to receive continuous updates via the NWDAF analytics subscription service or (2)~request specific analytics on demand. Each type of analytics is uniquely identified by a standardized analytics ID, as defined by 3GPP, and covers a diverse set of network aspects, including network performance (e.g., network function load), UE behavior (e.g., UE mobility and communication patterns), and security-related insights (e.g., UE abnormal behavior) \cite{3gpp_ts_23_288}.

Among these, UE abnormal behavior is a particularly significant analytics category, especially in the context of security automation. Its inclusion in the 3GPP specifications underscores NWDAF’s potential in enabling intelligent, zero-touch detection and mitigation of anomalous UE activities. 

According to 3GPP, abnormal behavior may be mobility-related, such as unexpected geographic location changes or unanticipated radio link failures, or communication-related, including potential Distributed Denial of Service (DDoS) attacks or excessively frequent service access attempts.

Each type of abnormal behavior is associated with a specific exception ID, allowing consumers to tailor their analytics requests by specifying either a general analytics category (e.g., mobility or communication) or a list of exception IDs. Based on the insights received, network functions can autonomously initiate predefined mitigation actions to address potential threats.

For example, in the case of suspected DDoS attacks, the SMF may release the PDU sessions of the malicious UEs and block them from establishing new sessions for a defined period. Similarly, the Policy Control Function (PCF) can also instruct the SMF to take such actions, reinforcing coordinated, policy-driven security enforcement across the network.~\cite{3gpp_ts_23_288}

\subsection{Literature Survey of NWDAF Research}

NWDAF has gained traction in recent years as a key enabler of intelligent, data-driven management in 5G networks. A few studies have explored various aspects of the function, with a focus on control plane monitoring, traffic analytics, and anomaly detection. In this section, we review key contributions in the literature, categorizing them based on their primary areas of focus.

\smallskip

\signpost{NWDAF for control-plane monitoring} 

Chouman et al. \cite{chouman2022towards} integrated a prototype NWDAF with the Open5GS 5G core network, establishing communication between NWDAF and other network functions via standardized interfaces—specifically, N34 (between NWDAF and the Network Slice Selection Function (NSSF)) and N23 (between NWDAF and the PCF). Notably, their implementation allowed NWDAF to access core network data even in the absence of explicitly defined interfaces with some functions. As part of their evaluation, the authors connected UERANSIM (a UE and RAN simulator) to the core network to collect control plane packets. They then analyzed these packets to extract statistics such as packet size and count per protocol. Furthermore, they performed a detailed analysis of the interaction between the Binding Support Function (BSF) and the Network Repository Function (NRF). A subsequent study by Manias et al. \cite{manias2022nwdaf} built upon the same experimental setup and dataset to further investigate interactions among network functions. By filtering packets in which both the source and destination were network functions, they compiled detailed statistics, such as average and maximum packet sizes per NF-to-NF communication path. Additionally, they applied k-Means clustering to categorize control plane traffic based on packet characteristics. However, these studies share a common limitation: they do not employ standardized event exposure services for data collection, nor do they support closed-loop automation. Instead, they rely on passive packet capture methods, which lack real-time interaction with NFs through event exposure services. As such, these implementations do not demonstrate NWDAF's full potential in enabling dynamic, automated network management.

\smallskip

\signpost{NWDAF for predictive analytics}
The NWDAF proposed in \cite{10588931} is integrated with the free5GC core and collects data from AMF and NRF via event exposure services. To supplement this, custom data collection methods were employed, including Google cAdvisor for resource utilization monitoring and a Python script to capture traffic between UEs and the Data Network Name. The NWDAF provides analytics on UE mobility (location), UE communication (uplink/downlink volume), and NF load (CPU, memory, and overall usage). These analytics are stored in Prometheus and visualized with Grafana. For predictive modeling, the authors used Linear Regression, Random Forest, and Decision Tree regressors trained on tcpdump-collected traffic data to estimate network performance. NWDAF performance was evaluated under single-user and ten-user scenarios to assess scalability and stability, running for over two minutes without failure. However, this implementation lacks UE communication data directly from the UPF event exposure service, which limits its accuracy and compliance with 3GPP standards. In~\cite{ibnbased}, the NWDAF is integrated with an Intent-Based Networking (IBN) platform for slice lifecycle management. Data is collected from the RAN, edge, and core using Grafana, Prometheus, and ElasticMon. The NWDAF uses ML-based predictions to inform IBN’s decision engine, enabling actions like resource scaling or DDoS mitigation to maintain QoS. However, it uses the OAI EPC, which lacks standardized 5G functions and does not leverage event exposure services, hindering its integration with other 5G cores.

\smallskip
\smallskip

\signpost{NWDAF for Anomaly Detection}
In \cite{cobra}, the authors implement NWDAF for anomaly detection in private 5G networks by extending the Open5GS core to support limited AMF and SMF event exposure (e.g., PDU session events and UES-IN-AREA-REPORT). A custom Traffic Monitor collects UE communication data, which is used to train models that detect anomalies such as low throughput and abnormal session patterns. However, the system does not respond to detected anomalies.

The NWDAF in \cite{mekrache2023combining} follows 3GPP specifications and adopts a microservice-based architecture. It gathers data from core NFs, VIM, and xApps, offering both core and ML-based analytics. Core analytics rely on data from AMF, SMF, UPF, and VIM, while the ML service uses an LSTM autoencoder to detect abnormal high-rate flows. Their SBI module, however, does not directly collect from UPF. Instead, traffic data is relayed from UPF to SMF via N4 and then to NWDAF through NSMF event exposure, an inefficient method, as direct UPF event exposure is supported by \cite{3gpp_ts_29_564}.

While these works align with 3GPP analytics and anomaly detection standards, they lack real-time closed-loop control and do not leverage UPF event exposure for UE traffic data.

It is clear that, as shown in Table~\ref{tab:nwdaf_summary}, leveraging NWDAF in closed-loop automation is limited in the literature. This gap, amongst others, is, however, addressed in our work.

\begin{table*}[ht]
\centering
\caption{Comparison of NWDAF Implementations}
\label{tab:nwdaf_summary}
\renewcommand{\arraystretch}{1.3}
\begin{tabularx}{\textwidth}{@{}L{2cm} L{2.5cm} L{1.5cm} L{1.5cm} L{1.5cm} Y@{}}
\toprule
\textbf{Work} & \textbf{Core Stack} & \textbf{Event Exposure} & \textbf{ML-based Analytics} & \textbf{Closed-Loop} & \textbf{Focus Area} \\
\midrule
\cite{chouman2022towards} & Open5GS & \ding{55} & \ding{55} & \ding{55} & Control-plane packet statistics \\
\cite{manias2022nwdaf} & Open5GS & \ding{55} & \ding{51} & \ding{55} & NF interaction clustering \\
\cite{10588931} & Free5GC & \ding{51} & \ding{51} & \ding{55} & Performance prediction (UE/NF) \\
\cite{ibnbased} & OAI EPC & \ding{55} & \ding{51} & \ding{51} & Slice scaling, DDoS mitigation \\
\cite{cobra} & Open5GS & \ding{51} & \ding{51} & \ding{55} & Abnormal UE throughput/sessions \\
\cite{mekrache2023combining} & OAI 5GC & \ding{51}\textsuperscript{*} & \ding{51} & \ding{55} & Flow-level anomaly detection \\
\textbf{This Work} & Open5GS \& OAI 5GC & \ding{51} & \ding{51} & \ding{51} & \textbf{Closed-Loop security (UE anomaly detection)} \\
\bottomrule
\end{tabularx}
\vspace{0.5em}

\raggedright
\footnotesize{\ding{51}: Supported \quad \ding{55}: Not Supported \quad *Uses SMF as relay for UPF data}
\end{table*}

\section{System Architecture and Design} \label{sec:system}
A self-managing network can be realized by integrating three critical steps: continuous monitoring, in-depth analysis, and dynamic action. However, these steps are still disjointed in most of the existing NWDAF solutions (See Table \ref{tab:nwdaf_summary}).
    
To address this, we propose a closed-loop NWDAF-enabled automation framework that unifies all three steps. First, we enable standardized continuous monitoring via the UPF event exposure service, enabling the collection of data from the user plane function in real time. Second, we enable in-depth analysis at the NWDAF, leveraging incoming telemetry data and dynamically provisioned ML models. Third, we extend the SMF to leverage NWDAF analytics to enforce security policies (e.g., session termination).

Section \S\ref{sec:system_overview} provides a high-level overview of our proposed system. In subsequent sections, we present details for the different components of our closed-loop system, namely, data collection using UPF event exposure (\S\ref{subsec:upf_event_exposure}), NWDAF analytics (\S\ref{subsec:analytics_architecture}), and SMF policy enforcement (\S\ref{subsec:smf_policy_enforcement}). 

\subsection{An overview of our system} \label{sec:system_overview}

\begin{figure}
    \centering
    \includegraphics[width=1.0\columnwidth]{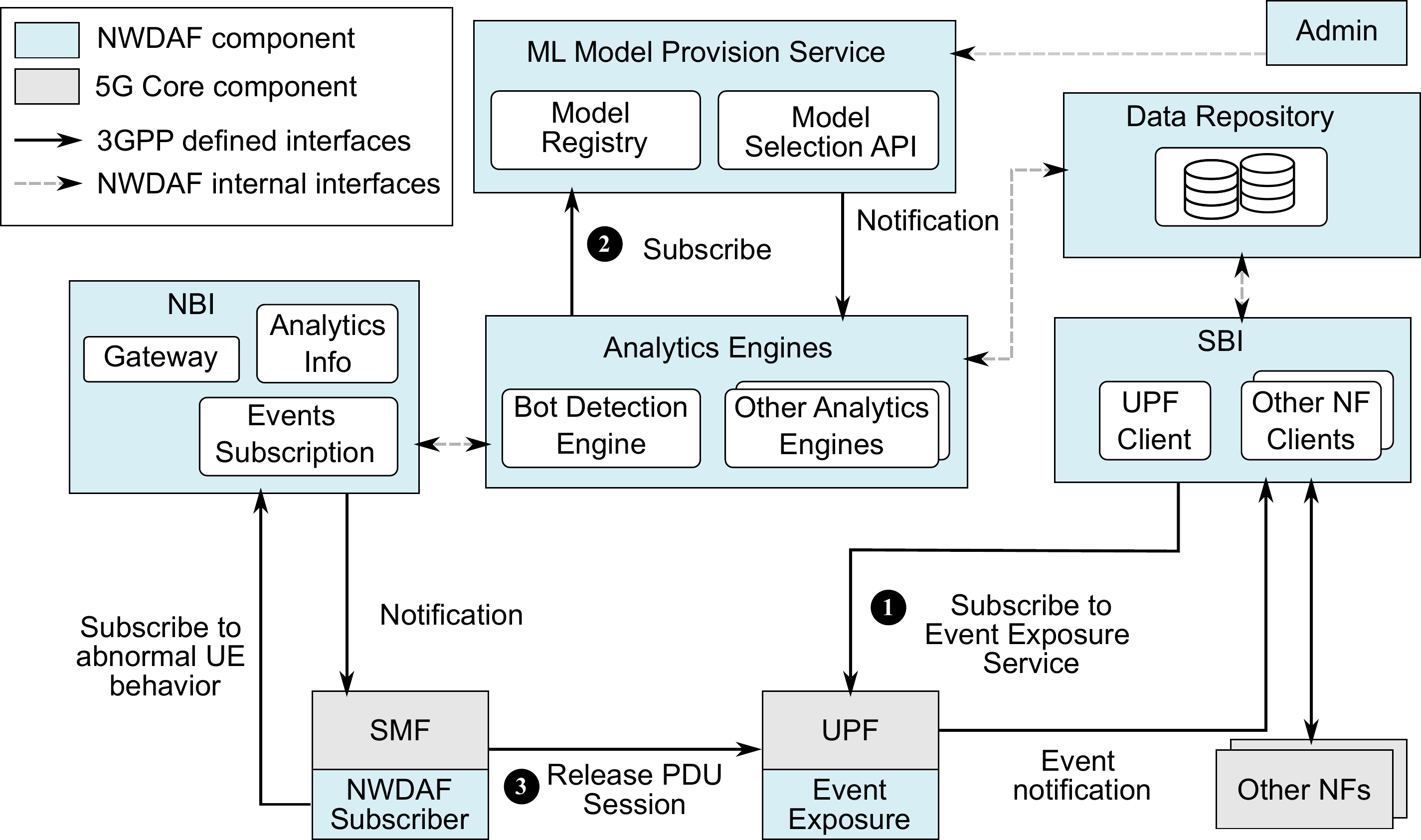}
    \caption{High-level overview of our NWDAF system for closed-loop management}
    \label{fig:system-overview}
\end{figure}

Our NWDAF system integrates continuous data collection, analysis, and action to realize closed-loop management. Figure \ref{fig:system-overview} shows a high-level overview of how these steps are realized within our system. We discuss each of these steps below.

\begin{itemize}[leftmargin=*]
\item \textbf{Continuous Data Collection}: The NWDAF south-bound interface (SBI) is responsible for collecting data from different sources (e.g., SMF, UPF). The UPF client inside this module sends a subscription request for data collection to the UPF event exposure service. In this request, NWDAF can indicate what data it is interested in collecting by specifying the standard fields like event type (e.g., \texttt{USER\allowbreak-DATA\allowbreak-USAGE\allowbreak-MEASURES}), measurement type (e.g., \texttt{VOLUME\allowbreak-MEASUREMENT}), or granularity of measurement (e.g., \texttt{PER\allowbreak-FLOW}). In addition, it can also indicate how it wants to receive data in the reporting mode fields, such as the frequency of receiving data from UPF (e.g., every second, every 10 seconds). 

After responding to the NWDAF's initial request, the UPF sends the data in the form of periodic reports to NWDAF, which then stores the data in its own database. Each of these reports has a timestamp field, allowing the NWDAF to learn about the historical behavior of the network in addition to monitoring its real-time status. This process is indicated as Step~1 in Figure \ref{fig:system-overview}.

\item \textbf{Data Analysis}:  NWDAF consists of different engines; each engine performs periodic analysis on the collected data in the NWDAF's database. This analysis can be performed by an ML model or statistical approaches.

If an engine wants to use an ML model, it must send a subscription request to the ML model provision service, specifying the necessary fields such as report type (e.g., \texttt{ABNORMAL\allowbreak-BEHAVIOR})  alongside some filters to help the service identify a suitable model. After accepting the engine's request, the provision service will send a URL to the engine, which can be used to run inferences on the ML model. Each engine will send its computed analytics to subscribers via the northbound interface. This is shown as Step~2 in Figure \ref{fig:system-overview}.

\item \textbf{Action}:
The 3GPP specifications \cite{3gpp_ts_23_288} provide potential actions that NWDAF subscribers can take based on reports of abnormal behavior. One of these suggested actions is for SMF to release the PDU session of the abnormal UE. This protects the network by banning abnormal UEs from sending traffic. We enable this action by extending the SMF to subscribe to the NWDAF. The SMF periodically checks NWDAF reports, and if an abnormal UE is reported, it releases the PDU session by sending a request to the UPF. This is shown as Step~3 in Figure \ref{fig:system-overview}, which completes the closed-loop management of the network.
\end{itemize}

\subsection{Data collection: UPF event exposure service} \label{subsec:upf_event_exposure}
\begin{figure}[!ht]
    \centering
    \includegraphics[width=0.9\columnwidth]{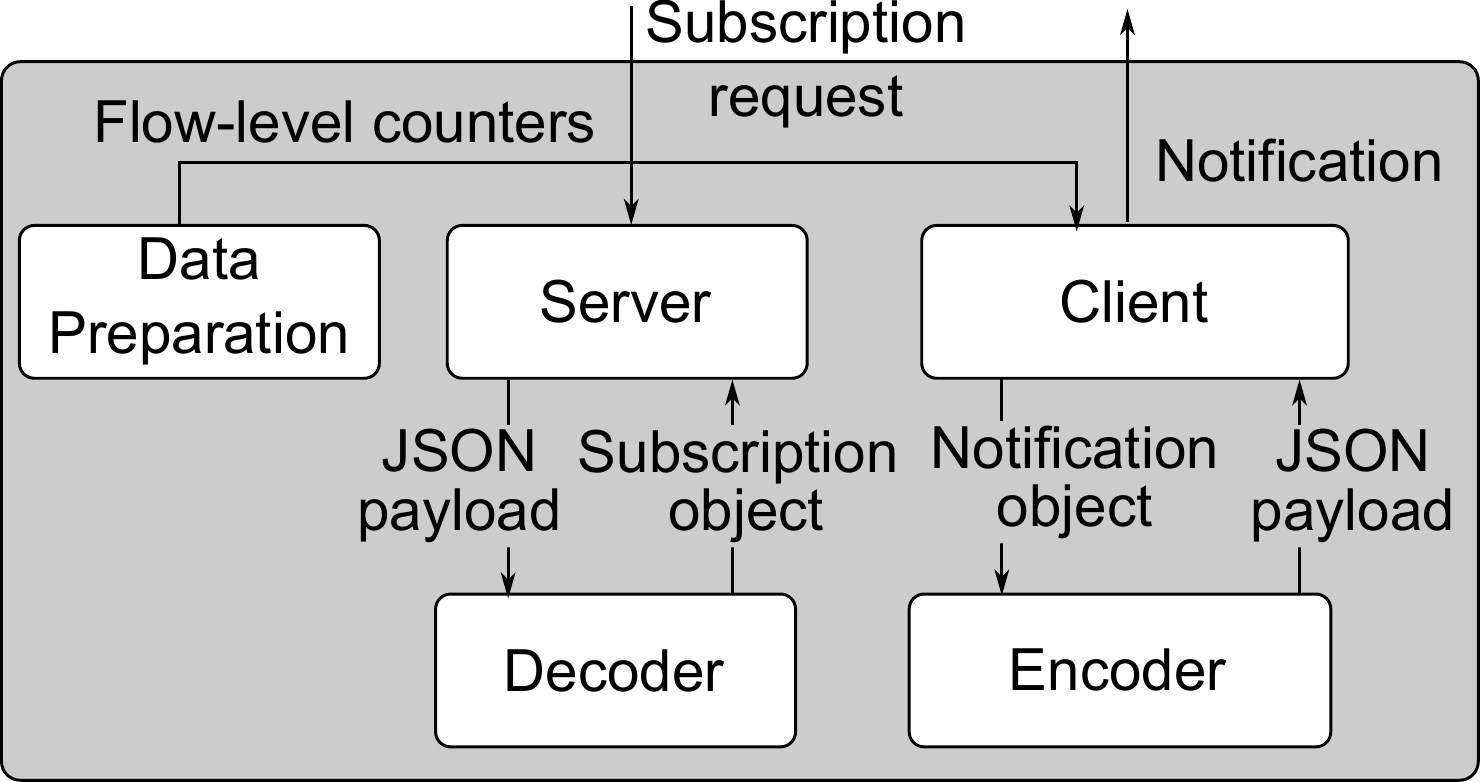}
    \caption{UPF Event Exposure Service}
    \label{fig:upf-details}
\end{figure}

We developed the UPF event exposure service by implementing five distinct modules, each responsible for handling a specific requirement of this service. Figure~\ref{fig:upf-details} shows these modules in the UPF. 

\signpost{Encoder/Decoder Module} All requests received by the event exposure service, as well as its corresponding responses, must comply with the format defined by the 3GPP specifications. To meet this requirement, we implement dedicated encoder/decoder modules with 3GPP standard format-compliant structured data types. Incoming requests to the Server will be parsed from JSON to these internal structures using the decoder, while outgoing responses and callbacks are encoded back into JSON using the encoder. 

\smallskip

\signpost{Server Module} The UPF event exposure service must handle different subscription requests, validate them, and send an appropriate response. In our design, the server module is responsible for these operations. It stores the subscription details, assigns a unique subscription ID, and sends a response back to the subscriber. The response includes the subscription ID and the details of the accepted subscription. If a subscriber requests events that are unavailable, the server omits them from the response. 

Since subscribers may ask for different kinds of data, it is essential to track each subscriber's requests when generating notifications. For this, we use a hashmap to associate each event with its corresponding subscription request. This approach allows an efficient lookup of subscribers interested in an event when sending notifications. As the subscriber list is shared with other modules to generate notifications, it is protected with a lock to ensure thread safety.

\smallskip

\signpost{Data Preparation Module} The data preparation module is designed to efficiently collect data without compromising performance. If there is at least one subscriber for this service, this module receives each packet passing through the UPF. Considering the high packet rate at the UPF, which significantly exceeds the notification generation rate, a lightweight approach is essential to collect data.

To achieve this, instead of performing different computations for each of the requested events and for each passing packet, we simply aggregate the critical statistics of each packet, such as packet volume and its direction. We use a hashing mechanism here to keep the collected data categorized by combining the flow identifiers (source IP, destination IP, source port, and destination port), the PDU session ID, and the subscription ID. This approach allows us to simply gather essential data during packet processing and defer the necessary computation to when we generate the notification. Our evaluation in the \S\ref{sec:evaluation} shows that this approach has limited overhead in terms of CPU usage and does not degrade the UPF throughput.

\smallskip

\begin{table}[htbp]
\centering
\caption{Measurements Supported by Our UPF EES}
\label{tab:supported-features}
\renewcommand{\arraystretch}{1.2}
\small
\begin{tabular}{@{}lp{0.55\linewidth}@{}}
\toprule
\textbf{Parameter} & \textbf{Supported Values} \\
\midrule
Event Type & \texttt{USER\_DATA\_USAGE\_MEASURES}, \texttt{USER\_DATA\_USAGE\_TRENDS} \\
Measurement Type & \texttt{VOLUME\_MEASUREMENT}, \texttt{THROUGHPUT\_MEASUREMENT} \\
Granularity & \texttt{PER\_FLOW}, \texttt{PER\_SESSION} \\
\bottomrule
\end{tabular}
\end{table}

\signpost{Client Module} The Client module is responsible for periodically generating and sending the appropriate notifications to each subscriber. To accomplish this, it compares the current time with the timestamp of the last sent report to determine if the reporting period indicated in the subscription request has passed. 

Depending on the subscriber's request, our implementation can generate various notifications. Table~\ref{tab:supported-features} lists the supported features. Each subscriber defines its expected notification by selecting the combination of these features. In this module, we implement different algorithms that generate various notifications from the basic data collected in the data preparation module.

Listing \ref{sample-notif} shows a per-flow notification item; 
Each of the measurement fields (e.g. \texttt{VOLUME\_MEASUREMENT}) in the notification reports includes different parameters. Each time a packet arrives, the data preparation module is triggered, while the server and client modules operate independently on separate threads. This configuration allows each module to operate simultaneously, preventing missing a subscription request or incoming packets.

\begin{lstlisting}[caption={Sample user data usage notification per flow}, label={sample-notif}]
{"eventType": "USER_DATA_USAGE_MEASURES",
  "ueIpv4Addr": "10.42.0.2",
  "snssai": {"sst": 2,"sd": "00002"},
  "timeStamp": "2025-03-27T18:03:49Z",
  "startTime": "2025-03-27T18:00:59Z",
  "userDataUsageMeasurements": [{
  "flowInfo": {
    "packFiltId": "{"SrcIp": "10.42.0.2", "DstIp": "142..}",
    "fDir": "BIDIRECTIONAL"},
    "volumeMeasurement": {...},
    "throughputMeasurement": {...},
    "throughputStatisticsMeasurement": {...}]}
\end{lstlisting}

\subsection{Analytics Architecture} \label{subsec:analytics_architecture}
To effectively achieve the NWDAF objectives, we adopt a modular architecture and build on OAI NWDAF design and implementation~ \cite{oai-cn5g}~\cite{mekrache2023combining}. In this architecture, there are separate modules for data collection (SBI), driving analytics (Analytics Engine), providing the ML models (ML Model Provision Service), and generating notifications (NBI). Figure~\ref{fig:system-overview} shows these modules inside the NWDAF. 

\smallskip

\signpost{Integration with UPF Event Exposure} To leverage the data provided by the UPF Event Exposure service, we implement a UPF client within the SBI module of NWDAF. Upon NWDAF deployment, the UPF Client will initiate the data collection process by subscribing to the UPF event exposure service and actively listening for notifications. Each received event will be processed using this module and stored in the NWDAF database. 

\smallskip

\signpost{Analytics Engine} Given that NWDAF is defined to support various types of analytics, each one can be handled by a dedicated analytics engine. These engines operate by retrieving the relevant data from the database, performing the analysis, and generating analytics output. This output will be packaged into the appropriate notification format by Event Subscription Service or Analytics Info to be sent to the consumer of NWDAF. 

NWDAF generates each analytic report using either statistical approaches or machine learning models. In addition, given the dynamic nature of network environments, it is essential to update and manage these ML models continuously. To address this requirement, we introduce the ML Model Provisioning Service. This service and its modules are shown in Figure \ref{fig:system-overview}. 

\begin{figure}[!ht]
    \centering
    \includegraphics[width=0.9\linewidth]{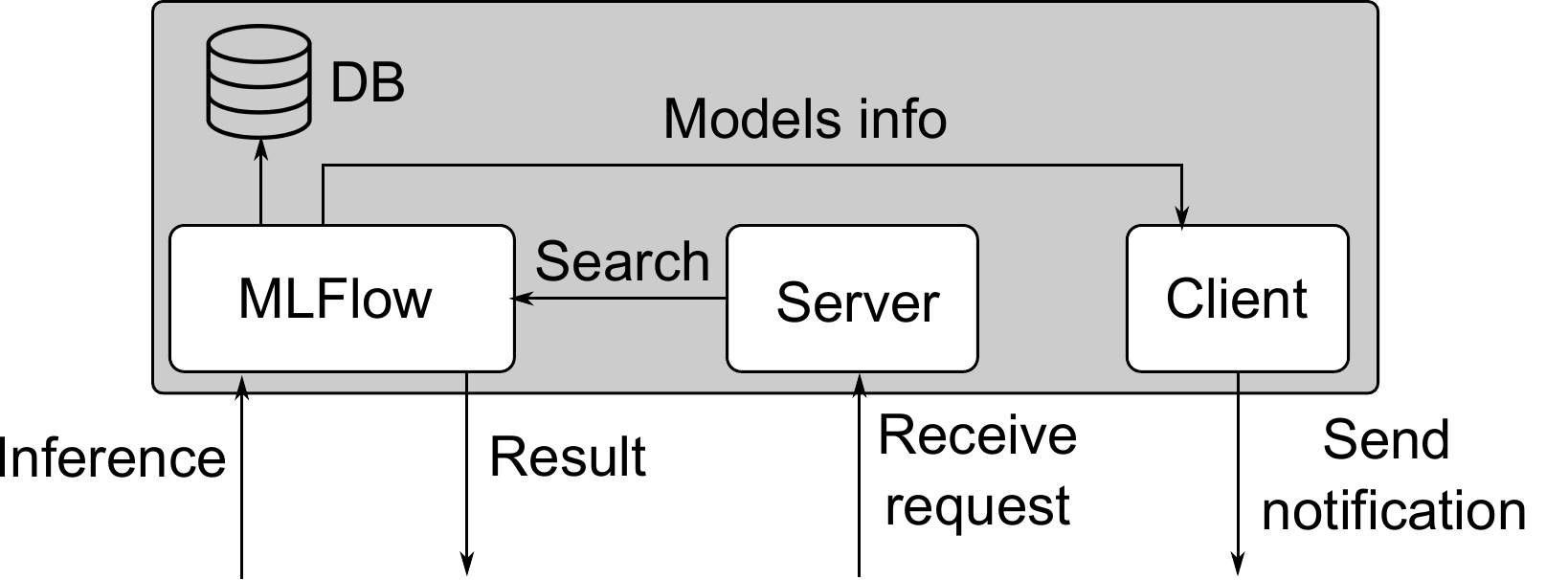}
    \caption{ML model Provision Service}
    \label{fig:ml-model-provision}
\end{figure}

\signpost{ML Model Provision Service} In line with the 3GPP specification, this service can receive different subscription requests through the server module and provide information about the requested ML models using the client module. We integrate MLflow \cite{mlflow-docs} as part of this service for model management and versioning. 
Each Analytics engine can subscribe to the ML model provision service, specifying the needed features for the requested ML model, following the format defined by the 3GPP specification. The service queries the MLflow registry to determine whether a suitable model is available. 

Considering the potential updates to the requested ML models, the service continuously monitors the MLflow model registry and notifies subscribers of the latest version of the ML models that match their request. This allows analytics engines to adopt the most recent and suitable version of the ML model that better matches their requirements.

\subsection{Automated Policy Enforcement via SMF} \label{subsec:smf_policy_enforcement}
While there are multiple actions that different network components can take after receiving the requested analytics from NWDAF, we focus on closing the loop by extending the SMF to act as an NWDAF subscriber.

\begin{table*}[t]
\caption{Summary of 3GPP Specifications vs. Our Contributions}
\label{tab:spec-vs-contrib}
\centering
\small
\begin{tabularx}{\textwidth}{@{} L{4cm} L{6cm} X @{}}
\toprule
\textbf{Component} & \textbf{3GPP Scope} & \textbf{Our Contribution} \\
\midrule
UPF Event Exposure & Specifies the format and semantics of event exposure messages & Implemented report generation and delivery with 3GPP-compliant encoding/decoding \\
NWDAF SBI & Suggests data collection using EES & Collects data from UPF using EES and stores it in MongoDB. \\
Analytics Engines & Defines report types and formats that NWDAF can generate & Implemented an ML-based Bot detection model with data processing and inference logic.\\
ML Model Provision Service & Specifies the format and semantics of providing ML Models & Implemented service logic and integrated with MLflow for model management \\
Analytics Subscriptions and Actions & Format for subscriptions, notifications, and action recommendations & Automated SMF extension for UE session release based on analytics \\
\bottomrule
 \end{tabularx}
\end{table*}

According to the 3GPP specifications, SMF can subscribe to the UE abnormal behavior reports provided by NWDAF. Upon receiving such a report, the SMF may initiate the termination of the PDU session of a suspicious UE, thereby preventing it from transmitting any further packets.

To enable this automated enforcement, we extend the SMF to subscribe to NWDAF, and receive periodic notifications. We add two modules\textemdash{} \textit{server} and \textit{client}, inside the SMF, where the client module sends a one-time subscription request to the NWDAF, and the server receives periodic notifications from the NWDAF's NBI module indicating if there is abnormal behavior in the network.

Subsequently, the SMF releases the PDU session of the suspicious UEs indicated in the NWDAF notifications. When an abnormal UE is detected, the SMF terminates the corresponding PDU session by instructing the UPF to release it via the N4 interface. This mechanism enables closed-loop mitigation of misbehaving UEs. Table \ref{tab:spec-vs-contrib} summarizes how our NWDAF implementation aligns with and extends the 3GPP-defined NWDAF capabilities.

\section{Implementation Details} \label{sec:implementation}

To realize our goal of 3GPP-compliant zero-touch network management, we integrated our NWDAF components with two widely used open-source 5G core network implementations \textemdash{} Open5GS \cite{open5gs} and OAI 5G Core \cite{oai-cn5g}. While the implementation details vary, both platforms support the entire closed-loop workflow consisting of real-time data collection, analysis, and dynamic action.

\smallskip

\signpost{UPF Integration} We implemented the Event Exposure service for Open5GS UPF in C and then integrated it with the existing Open5GS code base. The OAI-UPF-VPP component in the OAI 5G core patches a user plane gateway implementation using VPP. We implemented this in C, following the paradigm of the original code. We extended this implementation by adding the necessary modules to provide the UPF Event Exposure service. The Open5GS version supports a broader range of report types.

\smallskip

\signpost{NWDAF} In order to receive data from UPF, we implemented the UPF client in Go and inside the SBI component. The data collected by this module is stored in the MongoDB database. We developed the engines and the ML model provision service using Python. The ML Model Provision service also includes MLflow \cite{mlflow-docs}. MLflow provides a systematic framework for storing and managing various ML models, offering model version control and enabling a seamless pipeline for needed operations on ML models. We use the query option in MLFlow to efficiently search for models matching the filters specified in the request. 

Additionally, we enabled the custom API for the ML model registry. While there is an option for using the MLflow deployment terminal to register the trained models, we did that to make the registry of ML models easier. We also used the MLflow option to serve different ML Models for inference by receiving the inference data from the engines and returning model predictions. We modified other parts of NWDAF in Go to make the Engines reachable when generating reports for subscribers.

\smallskip

\signpost{SMF} We expanded the SMF in C to accommodate the Open5GS core. We also did the same in C++ to accommodate the OAI 5G core.
We examined how well our modules worked in \S\ref{sec:evaluation}; the results show that our Event Exposure Service adds negligible delay to the packets going through the UPF and can scale well with different numbers of UEs and changing data throughput. We also evaluated the detection latency of our closed-loop workflow; we show that data collection with shorter periods will lead to lower attack detection latency in our closed-loop system.

\section{Case Study: System in Action} \label{sec:case_study}

In this Section, we demonstrate the operational procedure of the proposed system by applying it to a real-time network management scenario involving abnormal behavior in the ongoing traffic of UEs. 

\smallskip

\signpost{Scenario Overview} To showcase our closed-loop NWDAF system's capabilities in a realistic and complex scenario, we focus on bot detection using the data collected from the UPF event exposure service. Detecting bot behavior requires more than per-UE throughput monitoring. This scenario showcases the real-time monitoring and analysis of the UE communication as well as the system's ability to mitigate abnormal behavior automatically. 

\smallskip

\signpost{Bot Detection Model} We utilize the bot behavior data from Scenarios 1, 2, 3, and 5 of the CTU-13 dataset \cite{ctu13} as our training dataset. We use the graph-based features to train a Random Forest model. We derive these features by creating a directed communication graph, treating each IP address as a unique node and assigning the number of transmitted packets as a weight to each edge. We extract the features of each node, including the number of incoming and outgoing edges, as well as the weighted indegree and outdegree of each node; we also include the weighted betweenness centrality of each node, showing how often a node appears on the shortest path between pairs of nodes in the network.  

As we expect our system to detect and mitigate the attacks while they are not finished, we should also train machine learning models on the data corresponding to the ongoing flows rather than complete flow or communication records. In our case, we considered the various destinations that the bot reached. For benign traffic,  we ran different scripts inside UEs and collected the data from NWDAF. 

\smallskip

\signpost{System Deployment} We deploy the system on Open5GS, which includes the modifications to the UPF and SMF mentioned in \S\ref{subsec:upf_event_exposure} and \S\ref{subsec:smf_policy_enforcement}. We use UERANSIM \cite{ueransim} to simulate gNB and UE behavior. After the deployment of the ML Model provision service, the trained model is registered using its admin API.

\begin{figure*}[!ht]
    \centering
    \includegraphics[width=0.65\textwidth]{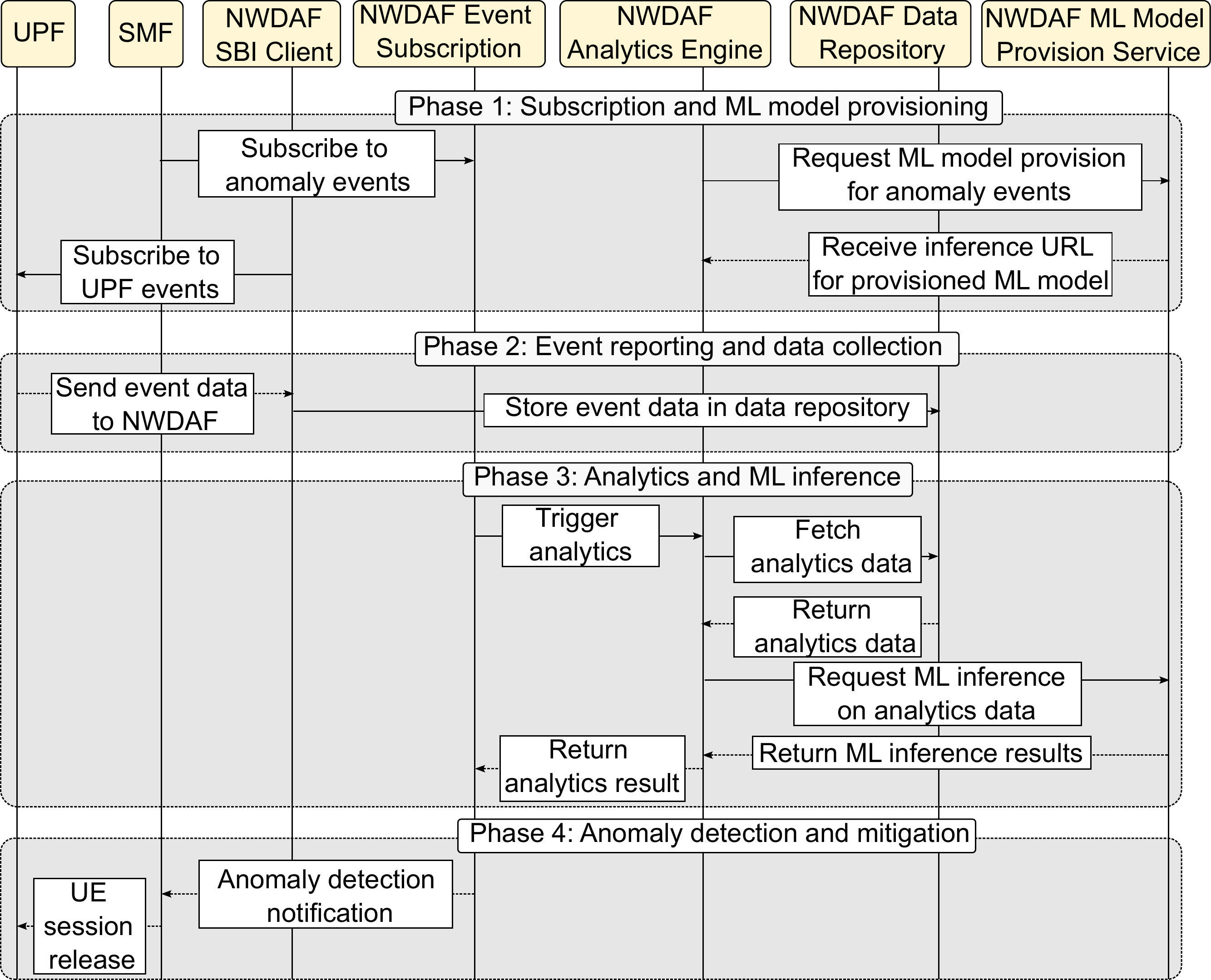}
    \caption{Sequence diagram for closed-loop workflow}
    \label{fig:sequence-diagram}
\end{figure*}

\signpost{Closed-loop Workflow} The sequence diagram \ref {fig:sequence-diagram} shows the different steps involved in the closed-loop workflow.

\begin{itemize}[leftmargin=*]
\item \textit{Subscription Phase}: Upon deployment, SBI subscribes to the UPF event exposure service, and the engine subscribes to the ML Model provision service, asking for a model for abnormal behavior events. Simultaneously, the SMF subscribes to NWDAF for the abnormal behavior detection event.

\item \textit{Data Collection Phase}: When the UEs start their communication, the UPF Event exposure service starts sending the periodic reports to the NWDAF SBI. 

\item \textit{Analysis Phase}: Based on the report period requested by SMF, NWDAF NBI will trigger the bot detection engine to investigate the network traffic. Bot detection engine then retrieves data from the NWDAF database and preprocesses it to derive the graph-based features of each UE communication; it will then use this and the inference model provided by the ML model provision service to investigate the behavior of each UE. 

\item \textit{Mitigation Phase}: The NBI receives the detected results from the bot detection engine and encapsulates them in 3GPP-compliant notifications and sends them to the SMF. If there is abnormal behavior reported in the notification, SMF will release the PDU session of the detected UEs to protect the network. 
\end{itemize}

\label{sec:evaluation}

\section{Performance Evaluation}
In this section, we evaluate the runtime performance and the resource usage of different modules of our system. All experiments were conducted using our NWDAF integrated with the Open5GS core, and using UERANSIM \cite{ueransim} for both UE and RAN simulation.

\subsection{UPF Overhead and Scalability} \label{subsec:upf_overhead}
We examine how well our developed event exposure service (EES) performs in terms of scalability and resource usage, as well as how it affects the routing capability of UPF. We compare three UPF variants:

\begin{itemize}[leftmargin=*]
\item \textit{Baseline}: Unmodified Open5GS UPF
\item \textit{EES : 0 Subscriber}: Our UPF EES implementation, but without active subscriptions.
\item \textit{EES: 1 Subscriber}: Our UPF EES implementation with one active subscriber, requesting per-flow volume reports every 3 seconds.
\end{itemize}

\begin{figure*}[!ht]
    \centering

    \begin{subfigure}[b]{0.32\textwidth}
        \includegraphics[width=\textwidth]{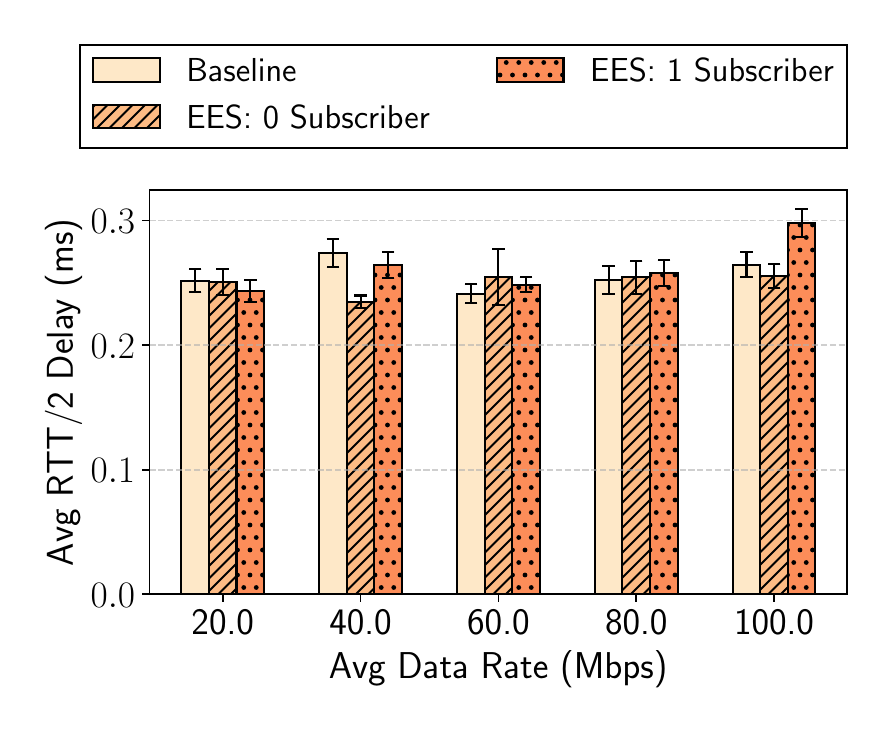}
        \caption{Latency remains unchanged with EES enabled, even under high data rates.}
        \label{fig:delay}
    \end{subfigure}
    \hfill
    \begin{subfigure}[b]{0.32\textwidth}
        \includegraphics[width=\textwidth]{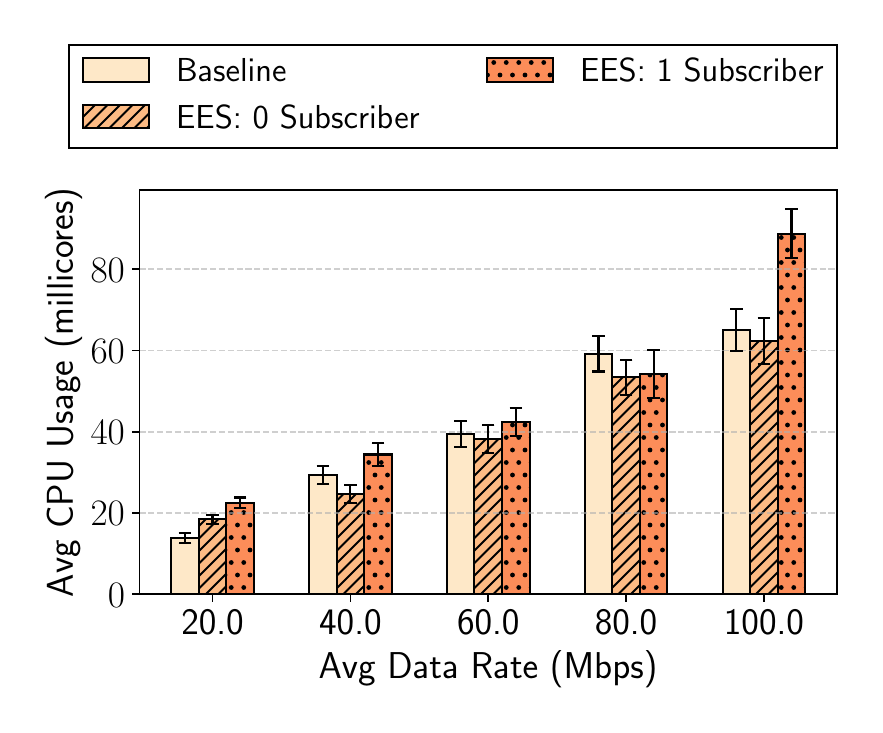}
        \caption{EES introduces modest CPU overhead; usage scales linearly with throughput.}
        \label{fig:upf_cpu}
    \end{subfigure}
    \hfill
    \begin{subfigure}[b]{0.30\textwidth}
        \includegraphics[width=\textwidth]{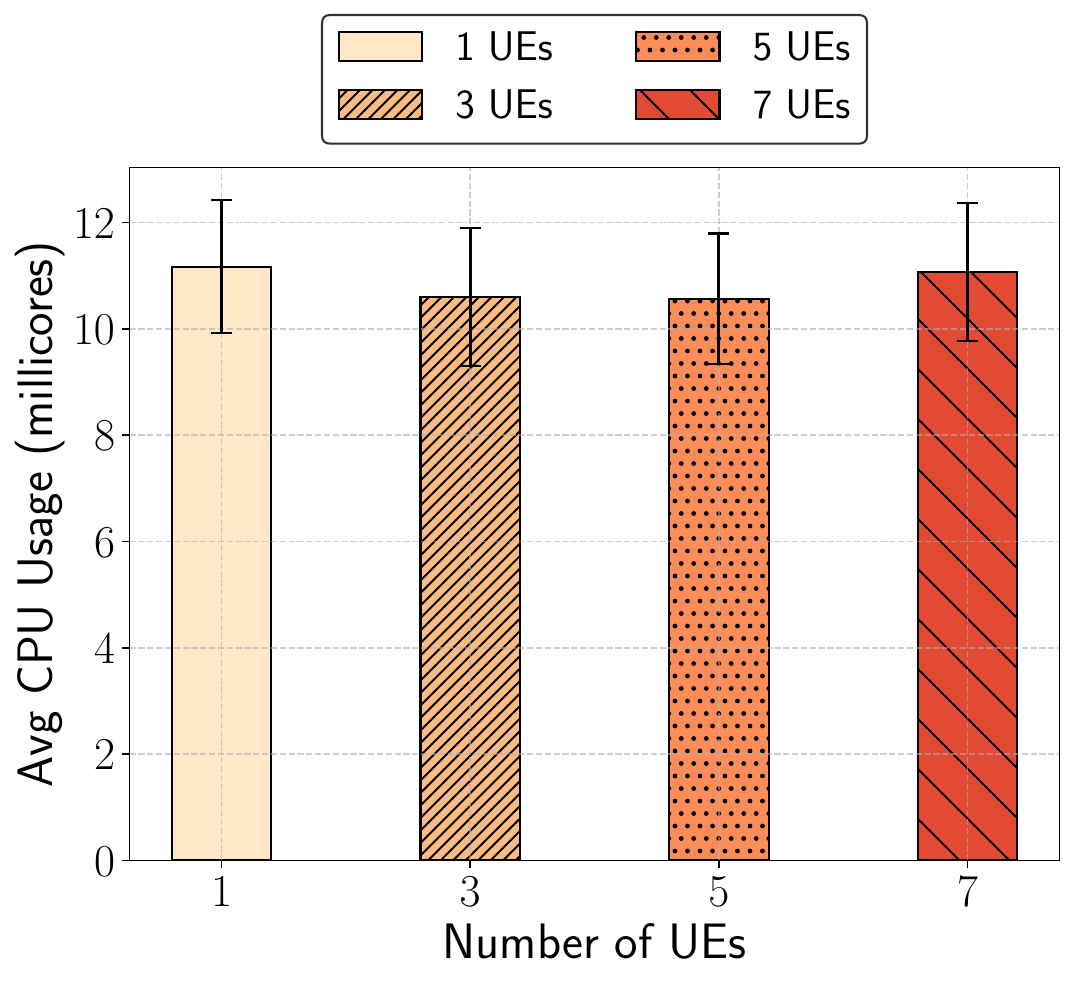}
        \caption{CPU usage remains stable as the number of UEs increases.}
        \label{fig:upf_vs_ue}
    \end{subfigure}

    \caption{UPF performance under varying stress conditions: (a–b) Increasing data rates; (c) Increasing number of UEs}
    \label{fig:upf-performance-combined}
\end{figure*}

Since the UPF EES performs real-time data collection and reporting, it may introduce additional CPU overhead or delay packet forwarding. We evaluate this overhead along two critical stress dimensions: (i) increasing data rate, which stresses per-packet processing load, and (ii) increasing number of UEs, which stresses the number of reporting flows the UPF must manage. Figure~\ref{fig:upf-performance-combined} presents the results with 95\% confidence intervals.

\smallskip

\signpost{Impact of data rate} To quantify the impact of EES on UPF performance, we compare the average one-way delay (RTT/2) and CPU usage of the UPF for the 3 UPF variants with varying data rates. For each given data rate, one UE runs iperf to another server in our testbed while another UE pings that server. We reported half of the average RTT of the ping results as latency for each experiment. Figures \ref{fig:delay} and \ref{fig:upf_cpu} show the outcome of this experiment. 

As the results show, there is no significant difference between the different versions of our UPF and the baseline in terms of latency, showing the minimal impact of our implementation on the performance of UPF, whose primary job is routing the packets. In addition, while there is no difference between the baseline and our UPF with no subscriber, there is a constant overhead added to our version with a subscriber due to the data collection and generating notifications. 
In our case, increasing the throughput means increasing the packets that should be processed for data collection. This demonstrates that our implementation scales effectively as throughput increases.

\smallskip

\signpost{Impact of number of UEs} Next, we examine the impact of the number of UEs on UPF performance. We fixed the total data rate at 10 Mbit/s but increased the number of UEs, resulting in having more traffic flows to report for UPF. Figure \ref{fig:upf_vs_ue} illustrates the results of this experiment. The result indicates that our implementation is scalable, and there is no significant change when increasing the number of UEs.

\subsection{NWDAF Resource Usage} \label{subsec:nwdaf_resource}

\begin{figure}[!ht]
    \centering
\includegraphics[width=0.9\columnwidth]{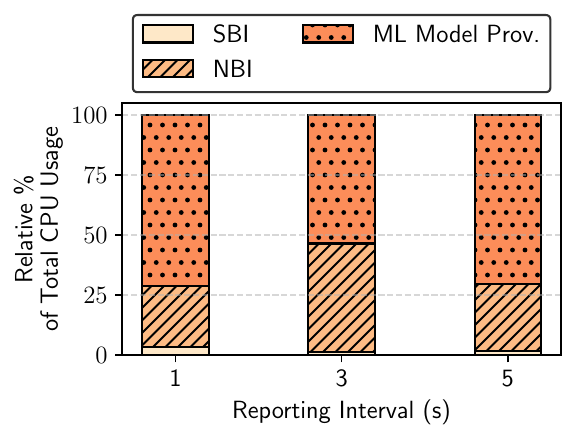}
    \caption{Resource usage of different NWDAF modules}
    \label{fig:nwdaf-res-usage}
\end{figure}

The NWDAF must support efficient real-time decision-making without incurring excessive overhead. To evaluate its scalability, we measure the CPU usage of its different components with varying reporting frequencies (1s to 5s). Figure~\ref{fig:nwdaf-res-usage} shows the relative percentage of CPU usage for each module. We see that the ML Model Provision Service makes up over 50\% of the CPU usage. This is expected since it performs compute-intensive tasks such as ML model querying, filtering, and inference. However, the other modules (e.g., SBI) use much less CPU.

\smallskip 

This highlights the fact that the data collection and mitigation aspects do not bottleneck the NWDAF and that the ML model provisioning should be independently scaled.

\subsection{Closed-Loop Workflow: Attack to Mitigation} \label{subsec:closed_loop}

To evaluate the responsiveness of our proposed closed-loop automation system, we examine the time taken for a bot attack to be detected and malicious UEs to be banned. 

\begin{figure}[!ht]
    \centering
\includegraphics[width=0.9\columnwidth]{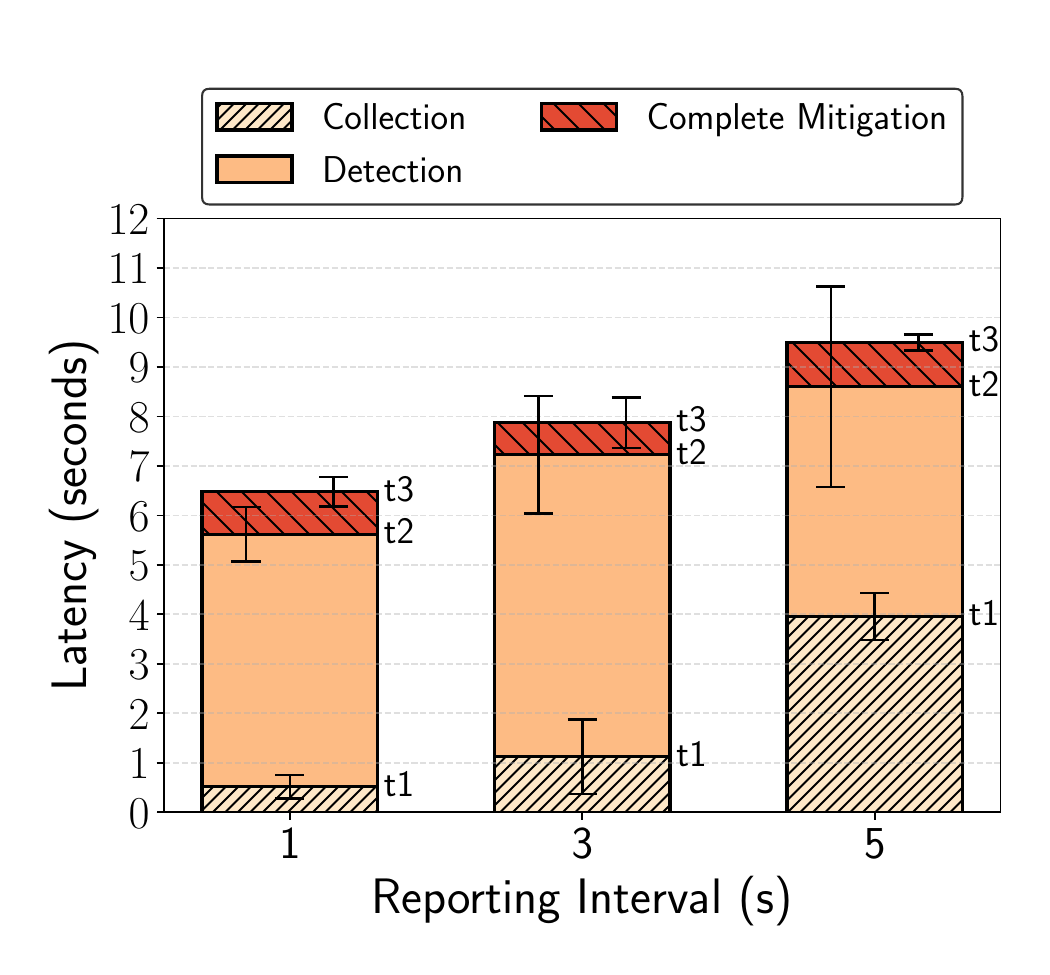}
    \caption{Response time breakdown for automated attack mitigation with NWDAF}
    \label{fig:latency_breakdown}
\end{figure}

We set up 20 servers with distinct IP addresses in our testbed to serve as the intended target of the bot attack. To simulate bot behavior, we used Nmap's \cite{nmap} --script http-open-proxy to scan for open web proxies. We configured SMF to receive abnormal behaviour reports from NWDAF every second, and varied the subscription period of NWDAF SBI for data collection from UPF. For each round of the experiment, after deploying the system, we started the bot behavior inside the UE. We also performed a ping inside the UE, setting a timeout of 0.5 seconds for responses. We used ping to illustrate the duration required for the network to ban the UE following the detection. For each data collection interval (1, 3, 5), we repeated this experiment 20 times and reported the mean and standard deviation. Figure \ref{fig:latency_breakdown} shows the result of this experiment.

In this figure, $t1$ shows the time taken after the start of the attack to see the first report of the UE malicious traffic on the NWDAF SBI. The average time of receiving the first report at the NWDAF in all cases is lower than the requested subscription periods.

Upon subscription, NWDAF sends a report of abnormal behavior every second to SMF; however, it takes until $t2$ for the NWDAF engine to have enough data for the detection of the bot pattern.
After the detection, SMF receives the notification from NWDAF and sends the request for the release of the PDU session to UPF through the N4 interface. It takes until $t3$ for UE to start getting a ping failure, showing that the UE has been completely banned from the network. 

While the mitigation time is almost the same in all three cases, we observe that shorter data collection intervals lead to faster mitigation. Additionally, the primary latency bottleneck in this pipeline is the time required for the ML model to detect abnormal traffic patterns in the collected data. This highlights an inherent trade-off: collecting data over a more extended period provides more information and can improve detection accuracy, but it also increases the response time. Optimizing this trade-off depends on the model’s characteristics and the required detection precision.

\section{Conclusion} \label{sec:conclusion}

This paper presented a 3GPP-compliant NWDAF-based closed-loop management automation framework for 5G networks. Our implementation includes the first realization of the UPF Event Exposure Service for standardized, real-time data collection from the user plane, an ML Model Provisioning Service integrated with MLflow for model lifecycle management, and enhancements to the SMF that enable it to act on NWDAF analytics.

We evaluated the proposed system using a concrete bot detection use case. We demonstrated that the UPF Event Exposure Service introduces minimal delay and scales effectively with increasing data rates and number of UEs. Our analysis of the response time of the closed-loop workflow showed that most modules are lightweight, with ML provisioning taking up most of the overall compute.

Our work demonstrates that incorporating NWDAF into the 5G control loop is both practical and efficient. The proposed system is extensible and applicable to other use cases related to anomaly detection and policy enforcement. 

Future work includes extending the pipeline to support online training and enabling adaptive ML models that evolve with network conditions.
\begin{acks}
This work was supported in part by funding from the Innovation for Defence Excellence and Security (IDEaS) program
from the Department of National Defence (DND).
\end{acks}

\bibliographystyle{ACM-Reference-Format}
\bibliography{main}

\end{document}